\documentclass{caosp306}

\usepackage{graphicx}
\usepackage{amsmath}
\usepackage{amssymb}
\usepackage{url}
\usepackage{natbib}
\articleNo{274}
\pubyear{2018}
\volume{48}
\volnumber{2}
\firstpage{329}
\received{January 26, 2018}
\accepted{April 3, 2018}

\begin{document}

\htitle{Some characteristics of the GLE on 10 September 2017}
\hauthor{V.\,Kurt, A.\,Belov, K.\,Kudela and B.\,Yushkov}

\title{Some characteristics of the GLE \\ on 10 September 2017}

\author{V.\,Kurt\inst{1}
         \and
        A.\,Belov\inst{2}
         \and
        K.\,Kudela\inst{3,4}
         \and
        B.\,Yushkov \inst{1}}

\institute{Skobeltsyn Institute of Nuclear Physics, Lomonosov Moscow State University, Moscow, 119991, Russian Federation, 
            \and
            IZMIRAN, Troitsk, Russian Federation, 
         \and
            Institute of Experimental Physics, Slovak Academy of Sciences, Ko\v{s}ice, Slovakia,\email{kkudela@saske.sk} 
         \and
            Nuclear Physics Institute of the CAS, \v{R}e\v{z}, Czech Republic}

%%%%%%%%%%%%%%%%%%%%%%%%%%%%%%%%%%%%%%%%%%%%%%%%%%%%%%%%%%%%%%%%%%%%%%%%%%%%%
%                        D A T E / R E C E I V E D
% Date inserted here will be the date when your paper was received The
% format is: month (not abbreviated), day, year.
%%%%%%%%%%%%%%%%%%%%%%%%%%%%%%%%%%%%%%%%%%%%%%%%%%%%%%%%%%%%%%%%%%%%%%%%%%%%%
\date{January 26, 2018}

\maketitle

\begin{abstract}
We present a short overview of the event associated with the recent
strong solar flare on 10 September 2017 (X8.2) based on the available
data both from satellite GOES-13 and from selected neutron
monitors. The onset time of SPE/GLE at 1\,AU was found between
16:06\,--\,16:08\,UT. The GLE effect was anisotropic with a maximum
increase of 6\%. The maximum energy of accelerated protons was
$\approx 6$\,GeV. We estimated the release time of sub-relativistic
protons into open field lines as 15:53\,--\,15:55\,UT.
\end{abstract}

\section{Introduction}

The flux of high-energy protons arriving at 1\,AU is associated with an energy release in a solar eruptive event and/or with the consecutive acceleration via a coronal mass ejection (CME). Solar Proton Events (SPE) or Ground Level Enhancements (GLE), are observed directly over a long time, most probably since the events on 28  February and 7 March 1942 when they were identified by \citet{Forbush1946} and named later as GLE 1 and 2, respectively. Reviews on solar proton events and on GLEs can be found, e.g., in papers by \citet{Shea1990} and \citet{Moraal2012}. The GLEs, which are important also for radiation dose at the airplane altitude, are analyzed according to data of a neutron monitor (NM) network \citep[e.g.,][]{Mishev2014}. Radiation hazard alerts are based also on the NM data if available in real time with high time resolution \citep[e.g.,][]{Souvatzoglou2014}.

Altogether, during the systematic investigation of GLEs, 72 events were recorded 
%\citep[see, e.g.,][the GLE database at the University of Oulu]{Belov2010, Poluianov2017}. 
(see, e.g., \citealp{Belov2010}; \citealp{Poluianov2017}; the GLE database at the University of Oulu). 
The real time database for high resolution neutron monitor measurements (NMDB) is accessible at \url{http://www.nmdb.eu} and described, e.g., by \citet{Mavromichalaki2011}.

A low-energy threshold of particles, detected by high-latitude neutron monitors, is $\approx 450$\,MeV (this threshold is specified by atmospheric absorption), but an effective energy exceeds 600\,--\,700\,MeV. Minimal detected energy for a medium and low-latitude NM is even higher; it is determined by the geomagnetic cutoff rigidity. There is no doubt that GLEs are connected with powerful solar eruptive events, but it is still debated whether protons responsible for the beginning of GLEs and high-energy SPEs are accelerated directly during a flare energy release or later when a shock wave propagates in the upper corona. Particle propagation in the interplanetary magnetic field (IMF) is a complex process controlled by a variety of factors. Angular separation of a site of observation (the Earth) and a source on the Sun affects this propagation \citep[e.g.,][]{Kallenrode1990,Tylka2006, Gopalswamy2013, Plotnikov2017}. In addition, the magnetospheric transmissivity has to be included to interpret correctly the ground based measurements. SPE observations up to proton energies of $\approx 700$\,MeV onboard GOES satellites allowed us to compare a time profile of each GLE obtained from data of the NM network with time profiles of high-energy proton fluxes observed in the outer magnetosphere where shielding by the geomagnetic field is very slight.

Here we present and discuss the recent GLE associated with a major eruptive event on 10 September 2017 that occurred in the active region NOAA\,12673 near the west solar limb (S05W88) with X8.2 importance (GOES-13).

\section{Data}

Let us consider the anisotropy at the beginning of the event. Usually the anisotropy is best clarified by comparison of count rates of northern and southern near-polar NMs. The asymptotic directions of NMs at high latitudes (not truly polar stations) have a rather narrow cone of acceptance in the longitude extent and they are collecting cosmic ray (CR) charged particles from regions near the equator. The ring of such stations is used for space weather studies \citep[Spaceship Earth;][]{Bieber95}. During a rather long time period, there was a pair of NMs looking towards north and towards south, namely Thule and McMurdo (by asymptotic directions indicated in Spaceship Earth). The NM installed at McMurdo has recently been moved by 200\,km and is now operating as a Korean Jang Bogo NM.

\begin{figure}[t]
\includegraphics[width=\textwidth]{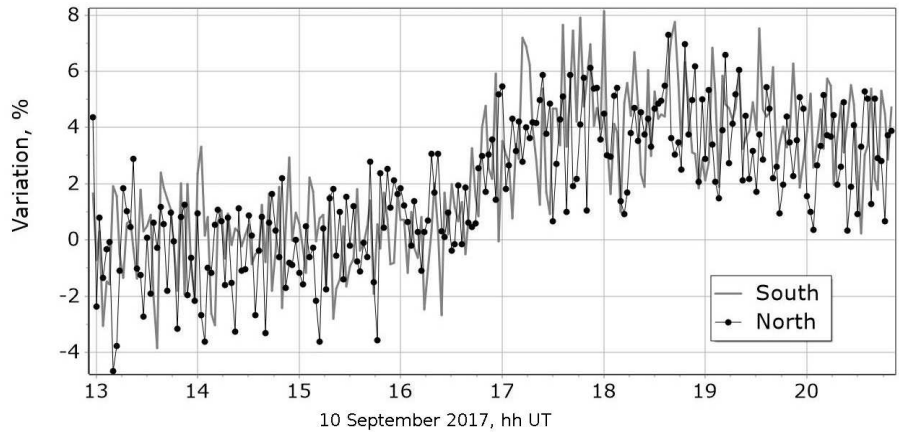}
\caption{The count rate variation of  northern (Thule) and southern (Jang Bogo) neutron monitors during the event on 10 September 2017. The variation is normalized to the average for one hour before the start of the GLE (14\,--\,15\,UT).}
\label{fig1}
\end{figure}

The comparison of the count rate of these polar NMs is presented in Figure~\ref{fig1}. This comparison did not reveal any north-south anisotropy. All high-latitude NMs situated at altitudes near the sea level have approximately the same dependence of the count rate on energy of primary CRs. They should display almost the same increases caused by solar CRs, if the effect is the isotropic one. There is a lot of such NM stations -- it is mostly an extended group of ground based CR detectors. However, just 11 suitable stations have been found in NMDB so far. Their nominal vertical geomagnetic cut-off rigidities $R_{\rm c}$ are $< 1.4$\,GV and  the standard atmospheric pressure is $>980$\,mbar.

\begin{figure}
\includegraphics[width=\textwidth,clip=]{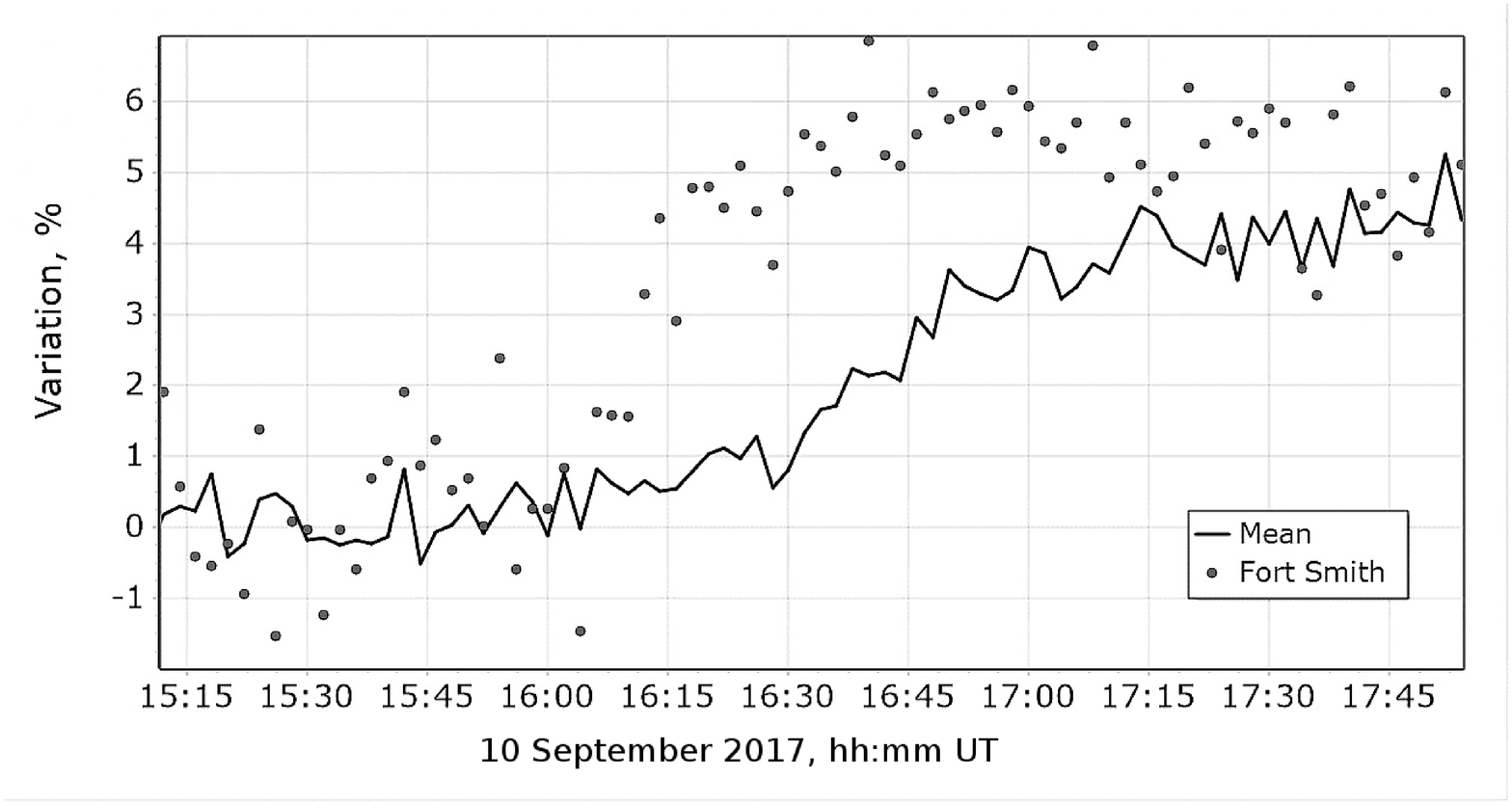}
\caption{Averaged variations of the count rate of high-latitude neutron monitors on 10 September 2017 (two minute averages, the smoothed line, Fort Smith is excluded) and variation at the NM Fort Smith (points).}
\label{fig2}
\end{figure}

We have averaged those "single-type" data from various NMs (see Figure~\ref{fig2}). For the averaged variation the statistical error is decreasing at least four times in comparison with 1-minute data of a typical NM. The main feature of the data is their rather high statistical error with the value above 1\%. The increase is well pronounced, however details are not easy to interpret. It was assumed that the 10 September 2017 GLE was isotropic from its very beginning. However, we can see that this was not the case. It is probable that it was a break through to the Earth of a very narrow stream of accelerated particles which was observed by a single NM Fort Smith (FSMT). Asymptotic cones of Thule, McMurdo (approximately the same for Jang Bogo) and of Fort Smith can be found in the paper by \citet{Kuwabara2006}. The FSMT NM, being one of stations of Spaceship Earth, has $<20^{\circ}$ extent of asymptotic longitudes and its asymptotic latitudes are close to the equator \citep[Figure~2 in][]{Kuwabara2006}. The fact that this NM is the only one which shows rather high increase among high-latitude stations indicates anisotropy of the GLE 72 in the first phase of its detection by NMs.

The different course of two temporal profiles of variations indicates the anisotropy of solar CR. The anisotropy was sufficiently high within the first hour of the event, which is typical for GLEs.

\begin{figure}
\centerline{\includegraphics[width=\textwidth]{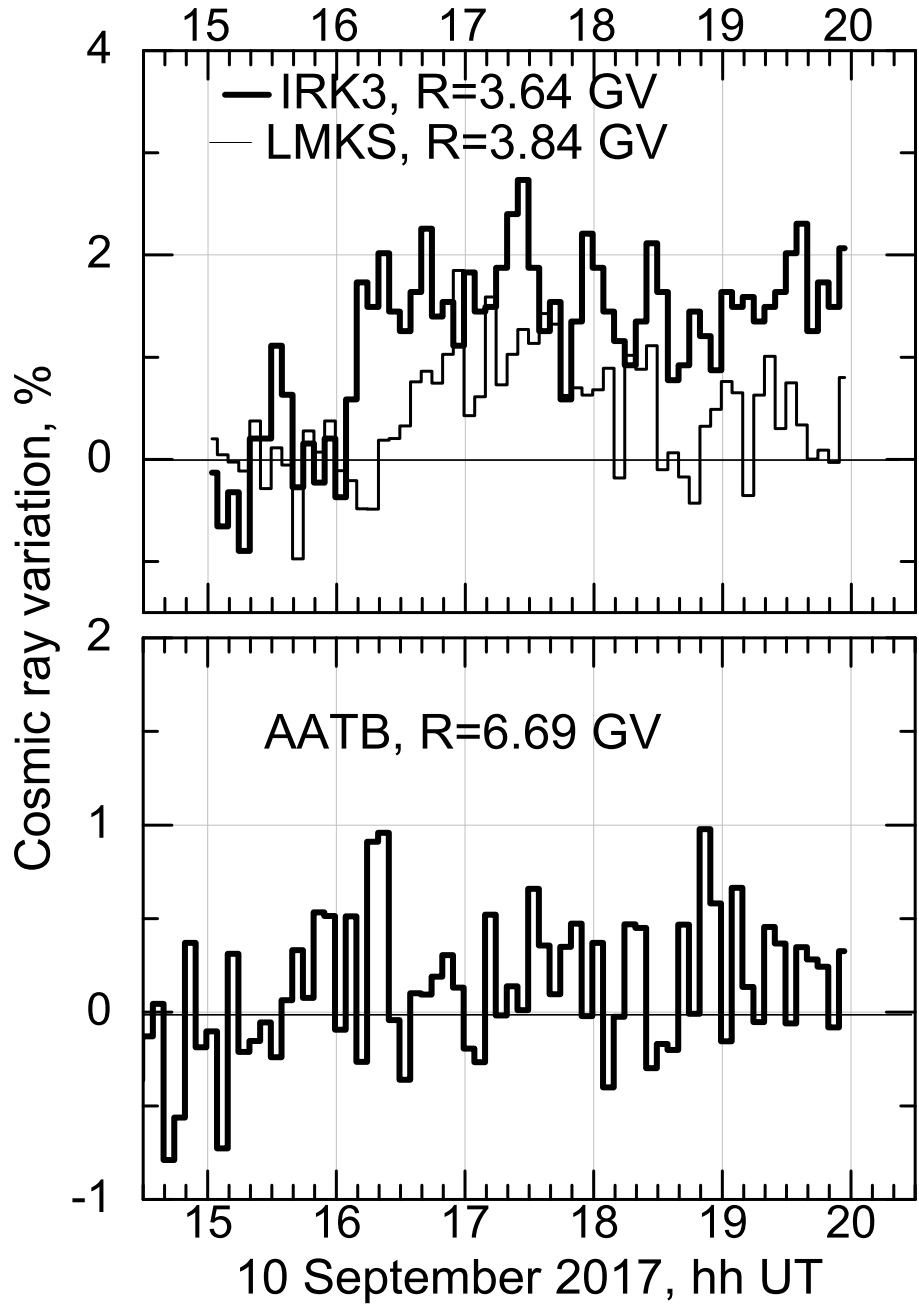}}
\caption{Variations of the count rate at selected middle-latitude stations.}
\label{fig3}
\end{figure}

Figure~\ref{fig3} displays time profiles of the count rate observed by three middle-latitude NMs. The count rate profiles of Irkutsk (IRK3) and Lomnick\'{y} \v{S}t\'{\i}t (LMKS) with almost similar cut-off rigidities situated at different longitudes (by $\approx 85^{\circ}$), indicate that anisotropy in the initial stage of GLE was clearly visible at higher energies. Although the
middle-latitude NMs have a rather large extent of asymptotic longitudes, the estimate of differences between the range of asymptotics of the IRK3 and of LMKS can be seen from Figure 3 (panels {\it a} and {\it d}) in \citet{Tezari2016}. While for IRK3 the spread of asymptotic longitudes is situated between $120^{\circ}$ and $\approx 260^{\circ}$, for LMKS it is situated between $\approx 40^{\circ}$ and $180^{\circ}$. Thus different time profiles of increases at two mid-latitude NMs probably indicate GLE 72 anisotropy which requires inclusion of more NMs and discussion of the pitch-angle distribution with respect to IMF. Here we selected two mid latitude NMs where some increase was observed from GLE 72. A small variation of the count rate of Almaty NM (AATB), where the cut-off rigidity is equal to 6.69\,GV, indicates that the maximum energy of accelerated protons most probably reached 6\,GeV.

\begin{figure}
\centerline{\includegraphics[width=\textwidth]{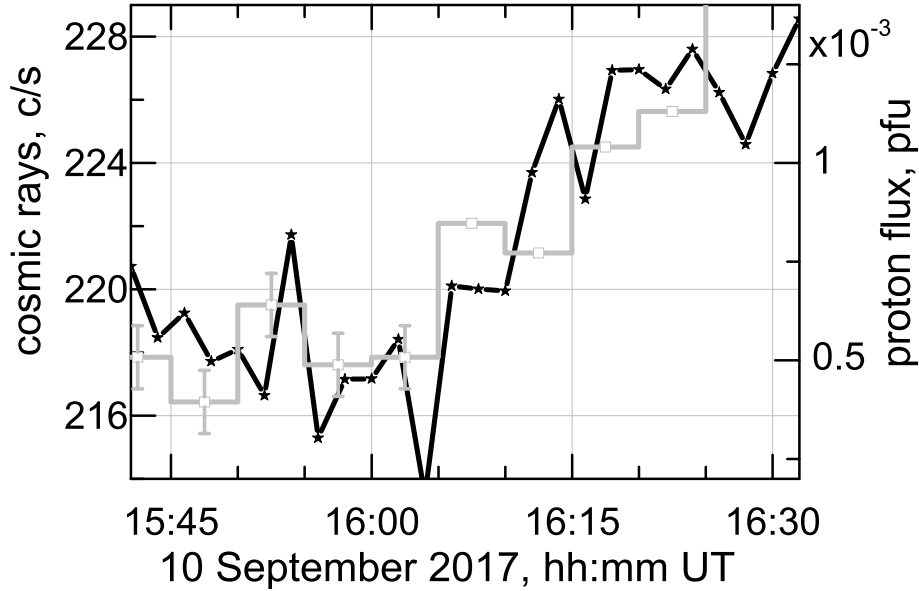}}
\caption{The count rate at NM Fort Smith (the black curve) and the flux of SPE protons with energies of 510\,--\,700\,MeV (data of GOES-13 -- the gray histogram).}
\label{fig4}
\end{figure}

Figure~\ref{fig4} presents the temporal variation at the NM where, unlike the other high latitude NMs, the increase was observed already at about 16:05\,--\,16:08\,UT. We compared this variation with the flux of solar cosmic rays in the 510\,--\,700\,MeV energy range (data of the GOES-13 satellite). GOES data indicate the onset time between 16:05\,--\,16:10\,UT. Thus we can say with confidence that the first SPE particles arrived to 1\,AU within the 16:06\,--\,16:08\,UT interval.

\section{Discussion and summary}

\begin{figure}
\centerline{\includegraphics[width=\textwidth]{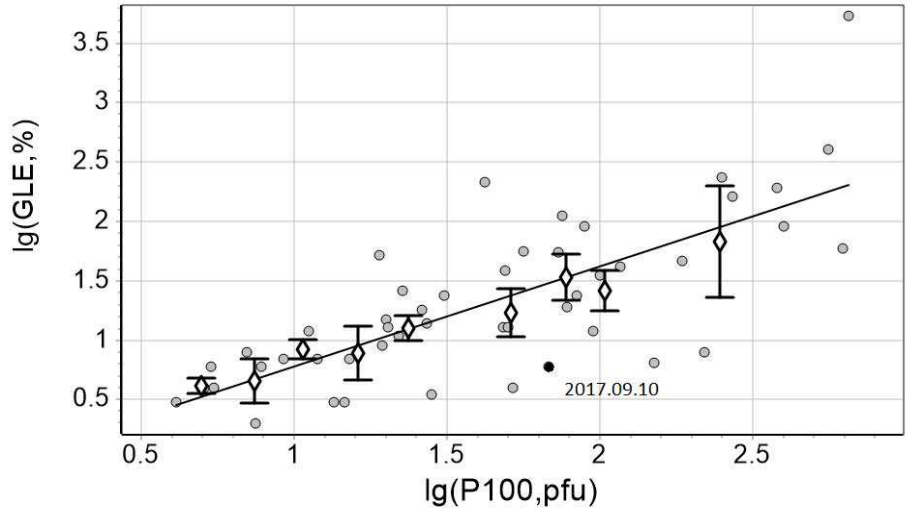}}
\caption{Dependence of the maximum increase of GLE on the maximum flux of protons with energy $>100$\,MeV for various GLEs. The black point corresponds to the GLE on 10 September 2017. The point in the upper right corner corresponds to the GLE on 20 January 2005. The linear fit is described by Equation~(\ref{eq1}).}
\label{fig5}
\end{figure}

The amplitude of this GLE associated with the 10 September 2017 (X8.2) flare which occurred at the western limb of the Sun, was relatively low -- a slight increase with approximately 6\,--\,7\%. Figure~\ref{fig5} presents a scatter plot of the maximum GLE increases observed on the ground since 30 April 1976 (GLE27) versus the maximum flux of SPE at the energy $>100$\,MeV (P100) observed on satellites (IMP and GOES data). The regression curve can be described as

\begin{equation}\label{eq1}
  \lg(GLE) = (-0.06\pm0.07) + (0.84\pm0.11)\lg(P100)\,.
\end{equation}

The event on 10 September 2017 is located within the diagram of the scatter plot, although relation of the GLE with respect to SPE is situated at the lower envelope of all events. This event is $5-6\sigma$ smaller than the average value indicated by curve (1). However, it is not the unique case among GLEs. There is a group of GLEs with rather soft energy spectra. Such types of spectra have been observed on 30 April 1976, 19 September 1977, 10 April 1981, 10 May 1981, 11 February 1992, 11 April 2001, and 17 January 2005 GLEs.

Maximum of the event was observed within 17:30\,--\,18:00\,UT interval. No north-south anisotropy was found in the event. However, at the beginning of the increase, a considerable longitudinal asymmetry was revealed, according to the difference between temporal behavior of the NM Fort Smith variation and the averaged variation of other high-latitude NMs  count rate. Given the different profiles of NMs IRK3 and LMKS (East-West), the longitudinal asymmetry is most probably imprinted at higher rigidities too, at least at $>3.6$\,GV.

It is supposed that GLE particles are accelerated at the front of the shock wave created in the solar corona during propagation of a CME \citep[see, e.g.,][]{Ryan2000, Yashiro2004, Kumari2017}. Another hypothesis is that the first high-energy protons arriving to the Earth's orbit are accelerated during the time when the essential energy amount in the flare was released. It is assumed that particle acceleration and subsequent plasma heating are closely connected with the energy release resulting from magnetic reconnection \citep[see,][]{Fletcher2011, Zharkova2011}. Both the process of the proton acceleration caused by the most intensive reconnection and the process of the acceleration caused by the shock wave should last for a certain period of time. We do not prefer either of these models.

Observations of the pion-decay emission during a solar eruptive event provide incontestable evidence of the proton acceleration up to high energies ($>300$\,MeV) and following interaction with the dense medium \citep{Ramaty1987, Vilmer2011}. When high-energy protons interact with matter, the pion-decay gamma-rays are emitted almost instantaneously. Fermi/LAT observed the onset of the high-energy emission ($E_{\gamma}>100$\,MeV) at $\approx$15:58\,UT (G.\,Share, private communication). This experimental fact means that the accelerated protons could not escape the Sun earlier than at 15:50\,UT.

On the other hand, given the observed time (16:06\,--\,16:08\,UT) of 600\,MeV SPE particles appearance at the Earth orbit, we can estimate the latest moment of particle release from the Sun. Suppose that these protons with $v\approxeq 0.8c$ propagated along the shortest IMF line $L \approx 1.2$\,AU. The propagation time is about 750 s, and consequently protons arrived to the Earth 250\,--\,300\,s later than any neutral emission. In other words, these particles escaped the Sun vicinity not later than 15:53\,--\,15:55\,UT.

It should be noted that in most events there are some uncertainties: a) of the time of recording the onset of acceleration by the observed onset of the high-energy gamma-emission $E_{\gamma}>100$\,MeV; b) exact knowledge of energies (velocities) of the protons responsible for the onset of the increase. These uncertainties allow to determine the escaping time interval of GLE particles into interplanetary space with several minutes accuracy. Let us consider the 20 January 2005 flare and GLE associated with it which has almost 100\% anisotropy and the amplitude of 6000\%. The first protons arrived to the Earth at 06:48:30\,UT\,$\pm$30\,s (above $5\sigma$ level). We compare this time with the time of appearance of the pion-decay emission, measured by CORONAS/SONG during the impulsive phase of the 20 January 2005 flare \citep{Grechnev2008, Masson2009, Kurt2013}. Even if the energy of the particles exceeded 10\,GeV, and they arrived to the Earth along the shortest possible trajectory $L \approx 1.1$\,AU, they had to leave the Sun at $\approx$ 06:38\,--\,06:39\,UT. The beginning of the pion-decay emission was observed by CORONAS/SONG at 06:44:40\,UT, and strong increase started from 06:45:30\,$\pm 5$\,s. Taking into account a photon propagation time we obtain that proton acceleration began within 06:36:40\,--\,06:37:40\,UT. Thus, even in this event, with  fairly accurate measurements of the onset of sub-relativistic proton acceleration on the Sun and the beginning of the GLE, it is possible to identify the time of particle release from the Sun with the accuracy of 3\,--\,4\,minutes.

\acknowledgements 
The authors wish to acknowledge Dr.\,G.\,Share for the high energy
gamma-ray data, PIs of all neutron monitors
(\url{http://www.nmdb.eu}), whose data are used in the paper, and GOES
data providers. KK wishes to acknowledge support of the grant agency
APVV project APVV-15-0194 and VEGA 2/0155/18. This paper was supported
by the project CRREAT (reg. CZ.02.1.01/0.0/0.0/15003/0000481) call
number 02\,15\,003 of the Operational Programme Research, Development
and Education. The authors are thankful to an anonymous reviewer and
the CAOSP editor for their help in improving the paper.

\end{document}